\documentclass[a4paper,twoside]{article}

\usepackage{xcolor}
\usepackage{hyperref}
\hypersetup{
    colorlinks,
    linkcolor={red!50!black},
    citecolor={blue!50!black},
    urlcolor={blue!80!black}
}
\usepackage{listings}	
\definecolor{KeywordBlue}{cmyk}{0.88,0.77,0,0} 
\definecolor{CommentGreen}{cmyk}{0.87,0.24,1.0,0.13} 
\newcommand{\spark}{SPARK 2014\xspace}
\newcommand{\gnat}{GNATprove\xspace}

\usepackage{multirow}	
\usepackage{booktabs}
\usepackage{longtable}

\usepackage{epsfig}
\usepackage{subcaption}
\usepackage{calc}
\usepackage{amssymb}
\usepackage{amstext}
\usepackage{amsmath}
\usepackage{amsthm}
\usepackage{multicol}
\usepackage{pslatex}
\usepackage{apalike}
\usepackage{algorithm2e}
\usepackage[bottom]{footmisc}

\usepackage{graphicx}

\usepackage{SCITEPRESS}     

\begin{document}

\title{Verifying LLM-Generated Code in the Context of \\Software Verification with Ada/SPARK}

\author{\authorname{Marcos Cramer\sup{1}\orcidAuthor{0000-0002-9461-1245} and Lucian McIntyre\sup{2}\orcidAuthor{0009-0008-0408-4340}}
\affiliation{\sup{1}secunet Security Networks AG, Dresden, Germany}
\affiliation{\sup{2}International Center for Computational Logic, Technical University Dresden, Germany}
\email{marcos.cramer@secunet.com, lucianmcintyre@mailbox.org}
}


\abstract{Large language models (LLMs) have demonstrated remarkable code generation capabilities, but the correctness of the generated code cannot be inherently trusted. This paper explores the feasibility of using formal software verification, specifically the SPARK framework for Ada, to ensure the reliability of LLM-generated code. We present Marmaragan, a tool that leverages an LLM in order to generate SPARK annotations for existing programs, enabling formal verification of the code. The tool is benchmarked on a curated set of SPARK programs, with annotations selectively removed to test specific capabilities. The performance of Marmaragan with GPT-4o on the benchmark is promising, with correct annotations having been generated for 50.7\% of the benchmark cases. The results establish a foundation for future work on combining the power of LLMs with the reliability of formal software verification.}

\onecolumn \maketitle \normalsize \setcounter{footnote}{0} \vfill

\section{\uppercase{Introduction}}
\label{sec:introduction}

Large language models (LLMs) have attracted significant attention within both the AI research community and the general public due to their generative capabilities. Tools such as ChatGPT have demonstrated the potential of LLMs to automate and accelerate processes in diverse fields, with software development benefiting particularly from their ability to generate code. However, while LLMs showcase impressive creativity and adaptability, they also present risks. As these models operate as black boxes, the code they generate cannot inherently be trusted to be correct or error-free, posing challenges for real-world applications where reliability and safety are essential.

To address the uncertainties associated with LLM-generated code, formal verification techniques offer a promising solution. Formal software verification employs rigorous mathematical methods to prove the correctness of code against a specified set of properties, which can help ensure that software meets its intended specifications reliably. Integrating formal verification with LLM-generated code has the potential to mitigate risks, making it possible to harness the creative benefits of LLMs while maintaining a high standard of code quality and safety.

The current paper is motivated by the need to bridge the gap between the creative potential of LLMs and the necessity for reliable, error-free code. By leveraging formal verification techniques, specifically through the SPARK programming language, we aim to explore the feasibility of combining LLMs with formal verification to produce code that is both innovative and provably correct. This study investigates whether LLMs can generate annotations for SPARK programs, facilitating formal verification of the resulting code. 

For this purpose, we have implemented Marmaragan, a tool that leverages an LLM in order to generate SPARK annotations for existing programs, enabling formal verification of the code using the GNATprove tool. Marmaragan can be viewed as a prototype for the backend of an AI-powered annotation generator that could run in the background of a SPARK editor. It incorporates features such as generating multiple solution attempts, retrying with additional context from GNATprove, and providing pre-compiled error messages to improve performance. Marmaragan can currently be combined with any LLM in the \mbox{OpenAI} API and has been most throughly tested with GPT-4o. It has parameters for the number of solutions generated in parallel, for the number of retries that Marmaragan attempts before giving up as well as for toggling a special chain-of-thought mode.

In order to evaluate how well Marmaragan performs depending on the value of its parameters, we created a benchmark based on a curated set of SPARK programs, from which annotations were selectively removed following various removal schemata. Experiments on the benchmark demonstrate Marmaragan's competence in generating annotations: Overall, it generates correct annotations for 50.7\% of the benchmark programs. Furthermore, the experiments shed light on what is the optimal balance between parallel solution attempts and retries in the light of limited computational resources. 

By successfully generating correct annotations, we establish a foundation for future work: In the near to medium-term future, this research could contribute to making applications of formal verification of code reliability and safety more efficient. In the long term it could contribute a building block towards a hybrid tool that combines the power of LLMs with the reliability of software verification for generating fully verified programs.

In Section~\ref{sec:preliminaries}, we discuss the preliminaries of this paper in the areas of logic, formal software verification (with a focus on SPARK~2014) and large language models. The implementation of Marmaragan is presented in Section~\ref{sec:implementation}. In Section~\ref{sec:benchmark}, we describe the methodology that we applied to create a benchmark for evaluating Marmaragan. The results of running Maramaragan with varying parameters on the benchmark are presented in Section~\ref{sec:results}, and in Section~\ref{sec:discussion} we discuss these results. Section~\ref{sec:related} presents related work. In section~\ref{sec:future-work}, we discuss future work before concluding in Section~\ref{sec:conclusion}.

\section{\uppercase{Preliminaries}}\label{sec:preliminaries}

This section discusses the foundations the work is set upon and the work it relates to and is inspired by. 

\subsection{Formal Software Verification}

Formal software verification is the process of proving the correctness of a software program with respect to a specified formal specification or property, using formal mathematical methods. It ensures that the software behaves as intended in all possible scenarios. In contrast with common non-formal testing techniques, which always cover only a limited number of scenarios and are thus vulnerable to having missed out on a scenario in which a bug takes effect, formal software verification covers all potential runs of the program. From now on, we will often use the equivalent term ``formal verification'' as a shorthand for ``formal software verification''.

There are different methodological approaches to formal verification. For this paper, we don't need to consider model checking and instead focus on deductive verification, which is ``the process of turning the correctness of a program into a mathematical statement and then proving it''~\cite{deductive-software-verificaiton}. In deductive verification, the desired behaviour of the program needs to be specified in a formal language. The task is then to prove that the program actually satisfies the specification for all possible inputs. 

At the level of single functions in the program, this is realized through pre- and postconditions, which are assertions on values of variables that enter and exit given functions within our program, specifying properties and relationships~\cite{hoare1969}. A precondition defines the conditions that must be met, so that a given function can be executed. Analogously, a postcondition defines the conditions that must be met directly subsequent to function execution. For example, a function computing $F(x,y) = x-y$ could have the precondition $x > y$ for ensuring that one stays in the realm of positive numbers. In this case, a sensible postcondition would be $F(x,y) > 0$, as this postcondition logically follows from the precondition and the definition of the function $F(x,y)$. This kind of logical entailment needs to hold for every postcondition of a function, and this needs to be established through a formal proof. We say that there is \emph{proof obligation} for deriving the postcondition.

\subsection{SPARK 2014}

SPARK~2014~\cite{spark2014} is a formally defined subset of the Ada programming language~\cite{ada}, designed to support the development of reliable and provably correct software~\cite{spark-book}. Its unambiguous semantics ensures that programs behave consistently and predictably. SPARK allows only specific constructs from Ada, ensuring compatibility with formal verification methods. Programs written in SPARK can be annotated with assertions, including \textit{preconditions} and \textit{postconditions}, to support modular deductive verification~\cite{hoare1969}. 

SPARK has found application in multiple areas, including train control systems and space transportation~\cite{spark-train-space}, commercial aviation~\cite{spark-aviation}, air traffic management~\cite{spark-air-traffic-management} and GPU design~\cite{nvidia}.

Some annotations in SPARK take the form of \texttt{pragma} statements, such as:

\begin{lstlisting}
pragma Assertion (condition);
\end{lstlisting}

These include constructs like \texttt{Assert}, \texttt{Loop\_Invariant} (see section~\ref{sec:loopinv} below), and \texttt{Loop\_Variant}, which facilitate detailed specification and verification. Additionally, SPARK ensures the absence of runtime errors, such as array bounds violations or division by zero, by verifying adherence to defined rules.

In SPARK, code is organized into two types of files: \texttt{.ads} and \texttt{.adb}. The \texttt{.ads} files, known as specification files, define the interface of modules, including function and procedure declarations along with their associated preconditions and postconditions. In contrast, the \texttt{.adb} files, or implementation files, contain the executable code and additional annotations such as \texttt{Assert}, \texttt{Loop\_Invariant}, and \texttt{Loop\_Variant} pragmas. This separation supports a clear distinction between the specification and implementation, facilitating modular reasoning and verification of SPARK programs.

The GNATprove toolchain is the primary mechanism for verifying SPARK programs. It operates in three stages:
\begin{itemize}
    \item \textbf{Check}: Ensures SPARK compatibility.
    \item \textbf{Flow}: Analyzes data and information flow.
    \item \textbf{Proof}: Verifies code against assertions and conditions using third-party theorem provers via Why3~\cite{why3}.
\end{itemize}

GNATprove translates SPARK code into proof obligations, resolving them using automated provers. This ensures compliance with user-defined and language-level constraints, making SPARK programs highly reliable.

GNATprove provides feedback in the form of \textit{errors} and \textit{mediums}. Errors typically indicate issues such as syntax or type errors that prevent the program from being executed. Mediums, on the other hand, result from proof obligations that could not be discharged, either because the statement being proved is false or due to missing annotations. When a statement is false, these mediums may include counterexamples generated by the tool to help identify the source of the~issue.

\subsubsection{Loop Invariants}
\label{sec:loopinv}

A \emph{loop invariant} is a property that holds during each loop iteration. It can be viewed as the induction hypothesis in an inductive proof over the number of loop iterations. Consider the example in Listing~\ref{lst:loop-invariant-example}.

\begin{lstlisting}[caption={Example of loop invariants for a SPARK function that doubles a number.}, label=lst:loop-invariant-example, float]
procedure Double_Number (X : in Natural; Result : out Natural) is
   Count : Natural := 0;
begin
   Result := 0;
   while Count < X loop
      pragma Loop_Invariant (Result = Count * 2);
      pragma Loop_Invariant (Count < X);
      Result := Result + 2;
      Count := Count + 1;
   end loop;
end Double_Number;
\end{lstlisting}

Here, the invariants state that the \texttt{Result} is twice the \texttt{Count} and that the loop counter does not exceed~\texttt{X}. 

\subsection{Large Language Models and Transformers}
The development of large language models (LLMs) has been a significant leap forward for AI development, spurred by the introduction of the transformer architecture by Vaswani et al. in ``Attention Is All You Need''~\cite{attention-is-all-you-need}.

LLMs, which are specialized neural models with billions of parameters, excel at capturing patterns in text data to perform a variety of language tasks. These models evolved from earlier statistical language models that relied on n-gram techniques to predict word sequences but struggled with long-range dependencies. Innovations like Word2Vec~\cite{word2vec} and GloVe~\cite{glove}, combined with the transformer's attention mechanisms, enabled the modeling of complex relationships between words (or rather between tokens), addressing the limitations of traditional approaches.

The transformer architecture, central to LLMs, employs a multi-layer encoder-decoder structure~\cite{attention-is-all-you-need}. Its core innovation, the self-attention mechanism, calculates the importance of tokens relative to one another, enhancing the model's ability to understand context. Multi-head attention further improves performance by allowing the model to focus on different aspects of input simultaneously. The final stages of transformers generate predictions through a combination of linear layers and softmax functions, transforming embeddings into meaningful output. These advancements, coupled with significant increases in model scale and training data, underpin the capabilities of state-of-the-art (as of 2024) LLMs like \mbox{OpenAI's} GPT-4 and GPT-4o models, which are the basis of this thesis.

\subsubsection{Chain-of-thought prompting}
\label{subsec:Chain-Of-Thought-Prompting}

LLM performance has been shown to depend on the formulation of the prompt that is given to the LLM. One prompting technique that is relevant to our work is chain-of-thought-prompting~\cite{Chain-of-Thought}, whose applicability to generating SPARK annotations we have studied (see Section~\ref{sec:implementation}). This prompting technique enhances reasoning by guiding LLMs through a series of intermediate natural language steps before generating the final output. For this, the prompt is extended by a note about the intended structure of the response, e.g.\ ``Let's think step by step''. This technique has been shown to significantly improve model performance on reasoning-heavy tasks, both for few-shot prompting~\cite{Chain-of-Thought} and for zero-shot prompting~\cite{zero-shot-CoT}.

\section{\uppercase{Implementation}}\label{sec:implementation}

This section provides a detailed description of Marmaragan,  a tool that leverages an LLM in order to generate SPARK annotations for existing programs, enabling formal verification of the code.

\subsection{Marmaragan}
Marmaragan \href{https://github.com/Elocien/Marmaragan}{[GitHub]} 
is a tool, developed in Python, designed for the \spark language. The tool implements a hybrid AI approach that combines the power of LLMs with the trustworthiness of logic-based reasoning to generate annotations required for formal verification within \spark programs. 

Use is made of the LangChain \cite{LangChain} Library to handle the calls to the OpenAI API and make LLM interaction seamless. 

Marmaragan takes as input an existing Spark project consisting of the specification and implementation files as well as any dependencies. Using this, it queries an LLM to generate missing $pragma$ statements and then allows \gnat to compile and verify the resulting code. The tool incorporates a range of strategies to assist in generating correct programs, including features for retrying with \gnat errors and mediums as well as post-processing of LLM output.

In the following, we describe the motivation and concepts behind Maramaragan. We survey the features of the tool and discuss each of the steps that are taken to transform input into output.

\subsection{Proof of Concept}
The aim was to develop a proof of concept for automatic annotation generation in SPARK. This concept stemmed from the hypothesis that generating formally verified code with an LLM circumvents the typical problems encountered with LLM-generated code.

Although LLMs are showing great ability in the area of code generation, the code they create cannot be assumed to be free of faults or bugs. By generating formally verifiable code, it is possible to eliminate these types of errors. Marmaragan is a step in this direction. By showing that it is possible to generate \spark annotations for a given program, we show that one of the main difficulties of generating formally verifiable code can be overcome. 

\subsection{Environment Emulation}
\label{subsec:environment-emulation}
Marmaragan is designed to emulate a tool which runs in the background of a \spark editor, such as \mbox{GNAT Studio} \cite{gnat-studio}. With this setup, a user may send requests to the tool, such that annotations are generated for their SPARK code.

As Marmaragan is a proof of concept, the idea was to design a tool which, given existing SPARK code, is capable of generating annotations. The resulting added annotations should lead to \gnat running free of errors and mediums. Given this premise, the prompting strategy and setup of Marmaragan is developed in such a way as to optimally emulate these conditions.

\subsection{Implementation Details}
\begin{figure*}[h]
    \centering
    \makebox[\textwidth]{\includegraphics[width=1.05\textwidth]{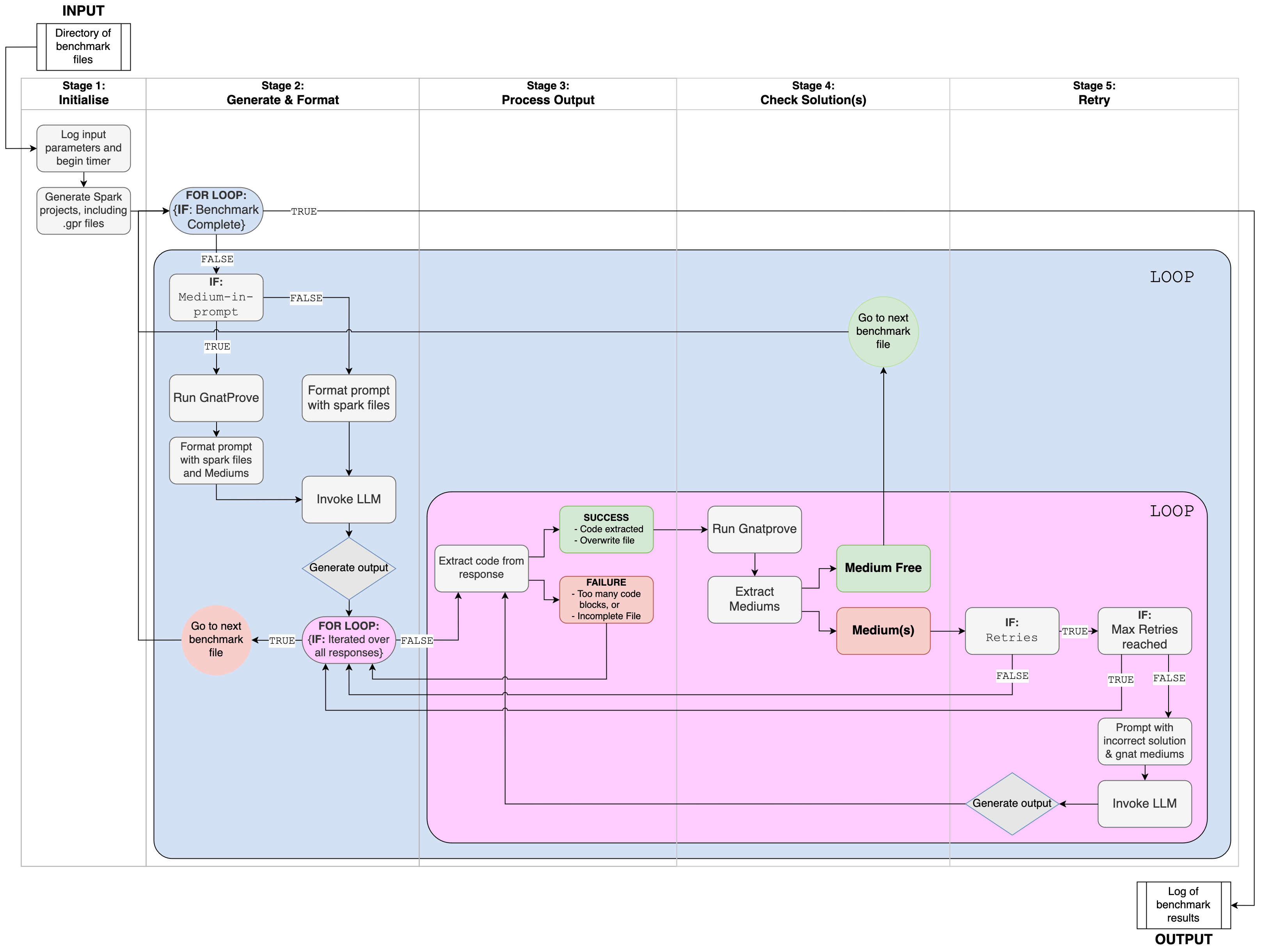}}
    \vspace{-1.6mm}
	\caption{Overview of the control flow in Marmaragan. Stage 1 initializes input parameters and generates the SPARK directory. Stage 2 begins the iteration over each of the files in the benchmark, formatting the prompt and invoking the LLM. Stage 3 is output processing, in Stage 4 \gnat is run and mediums are extracted. Processes in Stage 5 only run if enabled: the prompt is reformatted with the previous incorrect solution and \gnat medium(s) are also included.}
	\label{fig:marmaragan-process}
\end{figure*}

Here we deal with the implementation details of Marmaragan. Figure~\ref{fig:marmaragan-process} provides an overview 

\subsubsection{Marmaragan as a Benchmarking Tool}
Marmaragan is developed as a proof of concept, implemented as a benchmarking application in order to evaluate its functionality and performance. It works by taking a benchmark as input, then iterating over each of the files. For each task, it attempts to generate all required annotations, such that the code runs error- and medium-free. This procedure may be configured in multiple ways, by modifying the prompt to provide more context to the LLM and changing how solutions are generated.

\subsubsection{Prompting in Marmaragan}
As described in Figure~\ref{fig:marmaragan-process}, an important step of the workflow in Marmaragan is prompting the LLM. A good prompt is key to generating useful LLM responses, thus we delve into the details of this step.

In Marmaragan prompting works by inserting the given \spark files into the prompt, formatting and subsequently invoking the LLM. The prompt used for all queries can be found below in listing~\ref{lst:marmaragan-base-prompt}.

\begin{lstlisting}[caption={The base prompt used to query the LLM}, label=lst:marmaragan-base-prompt]
Try to solve the following problem logically and 
step by step. The final answer should then be  
delimited in the following way:

```ada

code here

```

The following are the specifications and 
dependencies of a Spark2014/ADA project:

{dependencies} 

This is the package body (implementation):

{package_body} 

Add one or multiple pragma statements 
(e.g. pragma Loop_Invariant, pragma Assert) to the 
package body, so that the code runs error and 
medium free.

Make use of the mediums provided in the prompt to 
guide your solution.

You must not modify the code in any other way, 
except to add "for" loops and "if" statements that  
enclose only pragma statements.

Do not modify the functionality in any way. Return 
the entire implementation file with the required  
additions.

\end{lstlisting}

The prompt is designed to be as general as possible, in order to adhere to the goal of emulation, as described in section~[\ref{subsec:environment-emulation}]. Therefore, no direct specification is given as to which $pragma$ statements to generate. 


It also includes the request that the code not be modified in any way, other than to add $pragma$ statements and possibly include structures such as ``for" loops and ``if" statements.

Additionally, when prompting LLMs with LangChain, it is possible to provide a \textit{system message}. The system message conveys to the LLM its context and which role it should take within the given context. As the system message for \mbox{OpenAI} GPT models is capped at 512 characters, the message chosen sticks to the key points the model should adhere to. The system message used is displayed below, in listing~[\ref{lst:marmaragan-system-message}].

\begin{lstlisting}[caption={The system message passed to the LLM}, label=lst:marmaragan-system-message]
You are a Spark2014/ADA programmer with 
strong logical reasoning abilities. 

You will be given an Implementation of a program, 
a specification of the program and the mediums 
that GnatProve raised for it.

You must complete the package body of the given
program, inserting one or multiple pragma 
statements.

You must not modify the code in any other way, 
except to add for loops and if statements that 
enclose only pragma statements, and do not 
modify the functionality.
\end{lstlisting}

\subsubsection{Medium in Prompt}
\begin{figure*}[th]
	\makebox[\textwidth][c]{\fbox{\includegraphics[width=1.1\textwidth]{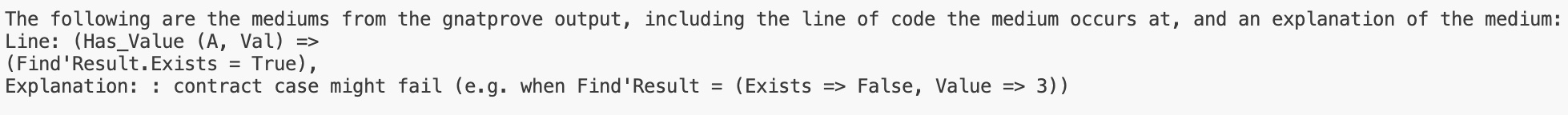}}}
	\caption[Pre-compiled GnatProve Mediums - Prompt Excerpt]{Example of Gnatprove precompiled and formatted medium messages, which are added to the prompt}
	\label{fig:medium-precompile}
\end{figure*}

Further, the prompt may be enhanced with additional context, by enabling the \textit{medium-in-prompt} feature. When this feature is enabled, \gnat compiles the SPARK project and any mediums from the output are extracted and formatted. In this case, formatting involves taking the line number and extracting the related line of code (and the one below) from the SPARK file. The line of code and the medium message is then appended to the prompt. Figure~\ref{fig:medium-precompile} gives an example of this. 

\subsubsection{Chain-of-Thought Prompt}
\label{subsec:natural-language-prompt}
As discussed in Section \ref{subsec:Chain-Of-Thought-Prompting}, chain-of-thought prompting \cite{Chain-of-Thought,zero-shot-CoT} is a strategy that helps to increase the performance of LLMs on complex reasoning tasks. To gain some insight into the effectiveness of different prompting techniques, a chain-of-thought prompt was developed, which deviates from the standard prompt only in the first section:

\vspace{0.1cm}

\begin{lstlisting}[caption={The beginning of Marmaragan's chain-of-thought prompt}, label=lst:marmaragan-natural-language-prompt]
Try to solve the following problem by first 
explaining in natural language what the underlying 
problem, leading to the medium, might be and how 
it could be solved. 
			...
\end{lstlisting}

Based on the results of the papers presented in \ref{subsec:Chain-Of-Thought-Prompting}, this prompt aims to make the LLM explain, in natural language, how it will go about proving the code. The hope is that this approach forces the model to reason more soundly. After explaining how it will solve the given program, it is then tasked with implementing this solution, all in one step, similar to the Zero-Shot Chain-of-Thought approach \cite{zero-shot-CoT}.

\subsubsection{N-Solutions}
The $N$-Solutions and $Retries$ parameters are fundamental instruments which can be employed to increase the benchmarking success rate. $N$-Solutions determines the number of responses an LLM returns, per individual prompt. At $ N=1$, the LLM supplies a single response to a given prompt, at $N=5$ it returns five responses. Due to how generation is affected by the \textit{temperature} parameter, each of the $N$-Solutions are in most cases distinct.

\subsubsection{Retries}
Setting the $Retries$ parameter to a value graeater than 0 makes Marmaragan continue with retries after a first failed attempt at generating pragmas that verify the code. The $Retries$ mechanism works by providing the LLM with additional context in the form of the previous failed attempt and the \textit{medium} messages generated by \gnat in that attempt. This additional context helps the model to formulate a new solution attempt. Increasing the number of $Retries$ leads to the additional solution attempts, each containing more context than the last.

\section{\uppercase{Benchmarking}}\label{sec:benchmark}
This section discusses the programs selected for benchmarking. This includes why the programs were chosen, where they were sourced from and how differing benchmarks were assembled from these. 

\subsection{Programs}
\label{subsec:programs}

In total, 16 \spark programs were selected, from three differing sources:

\begin{itemize}
\item Five programs originate from the \textbf{Argu Repository}, \href{https://github.com/marcoscramer/argu}{[Link]}\cite{argu}, 
which is a verified tool written by Marcos Cramer
for computing abstract argumentation semantics, with a focus on finding the grounded extension of an argumentation framework and proving its uniqueness. It was first published after the cutoff dates for the training of GPT-4 and GPT-4o, so that unlike for the other programs in the benchmark, we can be certain that it was not included in the training data of these LLMs. 

\item The \textbf{spark tutorial} \href{https://docs.adacore.com/spark2014-docs/html/ug/en/tutorial.html}{[Link]}, where the $linear\_search$ and $show\_map$ programs were taken from.

\item A repository of \spark implementations of common algorithms, known as\textbf{ spark-by-example} \href{https://github.com/tofgarion/spark-by-example}{[Link]}, where the final 11 programs were taken from, including basic implementations of $copy$ and $find$ algorithms, but also more complex programs such as $search\_lower\_bound$.
\end{itemize}

\subsection{Determining a Metric for Results Evaluation}
The choice was made to work with pre-existing, formally verified \spark projects, as this made the task of evaluating the results from the benchmarks possible. 

Quantifying how close a given solution is to being formally verified is very challenging. GnatProve provides no feedback in regards to this, excepting medium messages. These provide feedback about which statements lead to a failure in verification, but the total number of medium messages is not indicative of the closeness to a completed verification. A manual analysis is also not feasible, given the number of benchmark programs and the total number of solutions generated.

Thus, it is only possible to evaluate correctness by checking whether the program is completely free of errors and medium. By taking programs which are already verified, we are sure that a correct solution exists. Additionally, utilizing existing programs makes it possible to better curate which types of annotations to generate.

\subsection{Five Benchmarks}
In total, from the 16 programs, five benchmarks were developed with differing aims. For each benchmark, a separate schema for removing \textit{pragma} statements from the programs was devised. For some schemas, it was possible to do this multiple times per program. \textit{pragma} statements are removed only from the implementation file (\textit{.adb}) of the SPARK project. After removing pragmas from a program, we run GNATprove to check whether any mediums are generated. A program with removed pragmas is only included in the benchmark if GNATprove generates mediums for it, because otherwise it can be considered to be already fully verified, so that Marmaragan has no work to do on it.
\begin{figure}[ht]
	\centering
	\includegraphics[width=0.5\textwidth]{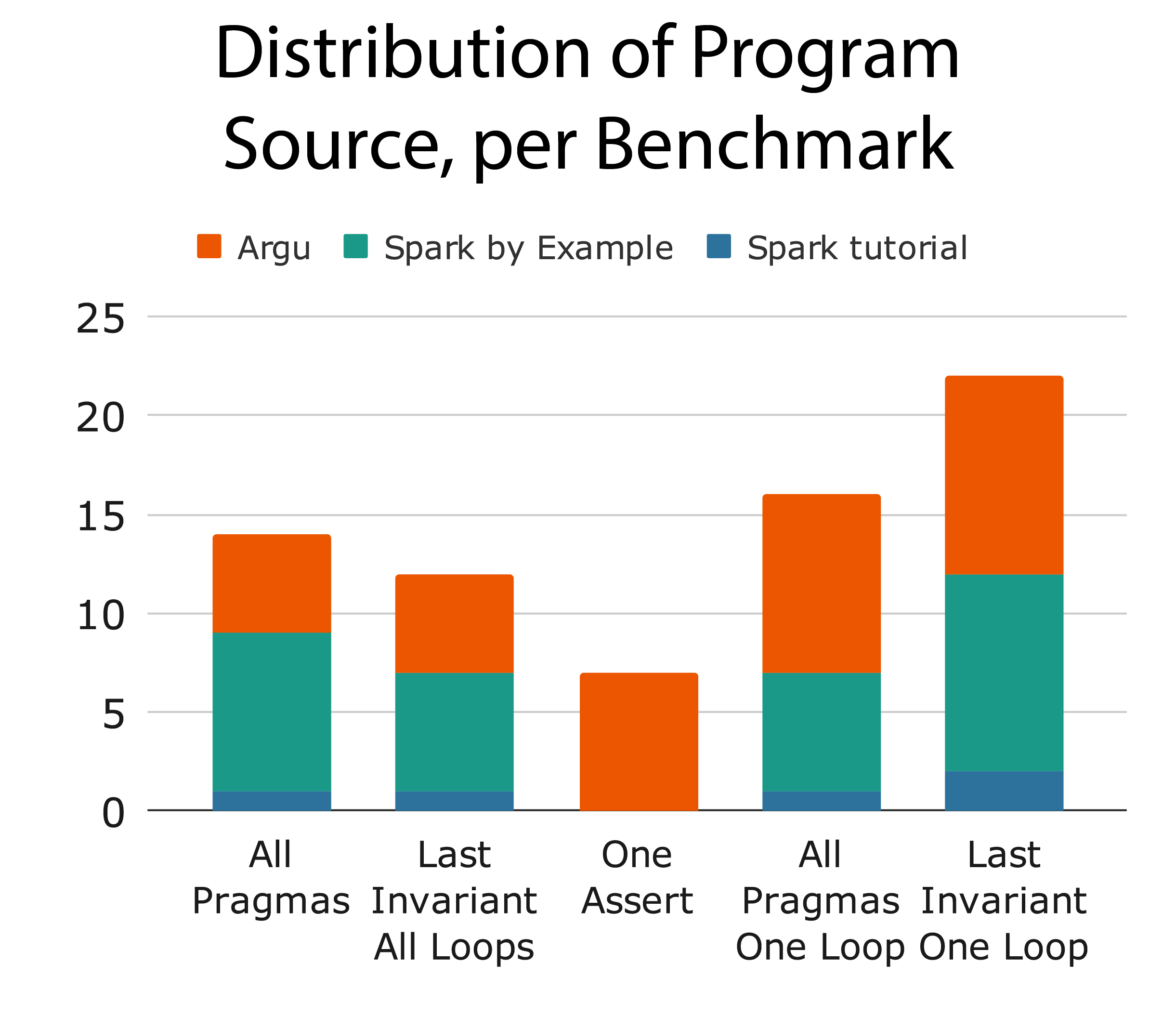}
	\caption[Bar Graph - Distribution of Source Programs in Benchmarks]{This chart details the distribution of programs from each of the three sources across the five benchmarks}
	\label{fig:distribution-graph}
\end{figure}

\hspace{-0.5cm}\textbf{All Pragmas}
\begin{itemize}
    \item All $pragma$ statements of the form $Loop\_\,Invariant$, $Loop\_\,Variant$, and $Assert$ are removed.
    \item 14 programs total
\end{itemize}
\textbf{Last Invariant All Loops}
\begin{itemize}
    \item The last $pragma\ Loop\_Invariant$ is removed from each loop in the file. If multiple loops occur in the same file, then multiple statements are removed.
    \item 12 programs total
\end{itemize}
\textbf{All Pragmas One Loop}
\begin{itemize}
    \item All $pragma$ statements are removed from a single loop, only for loops with two or more loop invariant statements. 
    \item 16 programs total
\end{itemize}
\textbf{Last Invariant One Loop}
\begin{itemize}
    \item The last $pragma\ Loop\_Invariant$ is removed from a single loop. 
    \item 22 programs total
\end{itemize}
\textbf{One Assert}
\begin{itemize}
    \item A single Assert is removed.
    \item 7 programs total
\end{itemize}

\vspace{0.2cm}
Figure [\ref{fig:distribution-graph}] displays a bar for each of the benchmarks, detailing the distribution of the source programs.

\section{\uppercase{Results}}\label{sec:results}
This section presents the findings from experiments with Marmaragan. The aim was to evaluate Marmaragan's performance across various benchmarks, while varying the individual parameters of the tool. Through a series of experiments, we attempt to measure the effectiveness of the tool and extract how the parameter $N$-Solutions and $Retries$ interact with each~other. 

\subsection{GPT-4o Release}
Early experiments conducted on the benchmark with vanilla GPT-4 demonstrated promising, albeit not entirely satisfying, results. However, shortly after these initial trials, GPT-4o was released. Not only did small scale experiments indicate that this new model was more successful overall, importantly, they also demonstrated that GPT-4o was far more cost-effective. This made larger experiments feasible. The main experiments were performed with the model \mbox{\texttt{gpt-4o-2024-05-13}}.

\begin{table*}[ht]
\centering
\scriptsize
\caption{The results of five experiments with unique combinations of $N$-$Solutions$ and $retries$, over five different benchmarks. The final column displays the total number of programs solved across all benchmarks.}
\begin{tabular}{lcccccc} 
\toprule
          & \multicolumn{1}{c}{All pragmas} & \multicolumn{1}{c}{Last invariant all loops} & \multicolumn{1}{c}{One assert} & \multicolumn{1}{c}{All pragmas one loop} & \multicolumn{1}{c}{Last invariant one loop} & \multicolumn{1}{c}{Sum}\\ 
\cmidrule(lr){2-2}\cmidrule(lr){3-3}\cmidrule(lr){4-4}\cmidrule(lr){5-5}\cmidrule(lr){6-6}\cmidrule(lr){7-7}
n=12, r=0 & 3 & 4 & 5 & 2 & 6 & 20 \\
n=6, r=1  & 2 & 3 & 6 & 1 & 12 & 24 \\
n=4, r=2  & 3 & 4 & 7 & 3 & 7 & 24 \\
n=3, r=3  & 2 & 4 & 7 & 0 & 7 & 20 \\
n=2, r=5  & 2 & 4 & 6 & 0 & 6 & 18 \\
\bottomrule
\end{tabular}
\label{tab:major-experiments}
\end{table*}

\subsection{Experiment Setup}
Adjusting for these new possibilities, a large scale experiment to test the capabilities of Marmaragan was devised. The aim was to derive the effectiveness of the \textit{$N$-Solutions} and \textit{Retries} parameters. Various combinations of each of the parameters were conceived to test this. A central goal of these tests was to determine what balance between $n$ (\textit{$N$-Solutions}) and $r$ (\textit{Retries}) was ideal in order to get the best results given a fixed amount of computational resources available for completing the verification of a program.

Given the values $n$ and $r$ for the \textit{$N$-Solutions} and \textit{Retries} parameters, the number of solutions generated by the program is limited to a maximum of $n(r+1)$, because $n$ solutions are generated for the first attempt and $n$ further ones for each of the $r$ retries. The total number of solutions to be generated per benchmark program was set to 12, as this made various combinations of $n$ and $r$ possible, while keeping costs within the limits set. The resulting combinations were the following:
\begin{lstlisting}
(n, r) combinations:
   (12, 0),   (6, 1),   (4, 2),   (3, 3),   (2, 5)
\end{lstlisting}

\subsection{Experiment Results}
Table \ref{tab:major-experiments} details the results of each of the five experiments on each of the five benchmarks. 

In total 36 programs out of 71 were solved across all five benchmarks, meaning a solution was found for 50.7\% of all benchmark problems. 

\begin{figure}[h]
	\centering
	\includegraphics[width=0.5\textwidth]{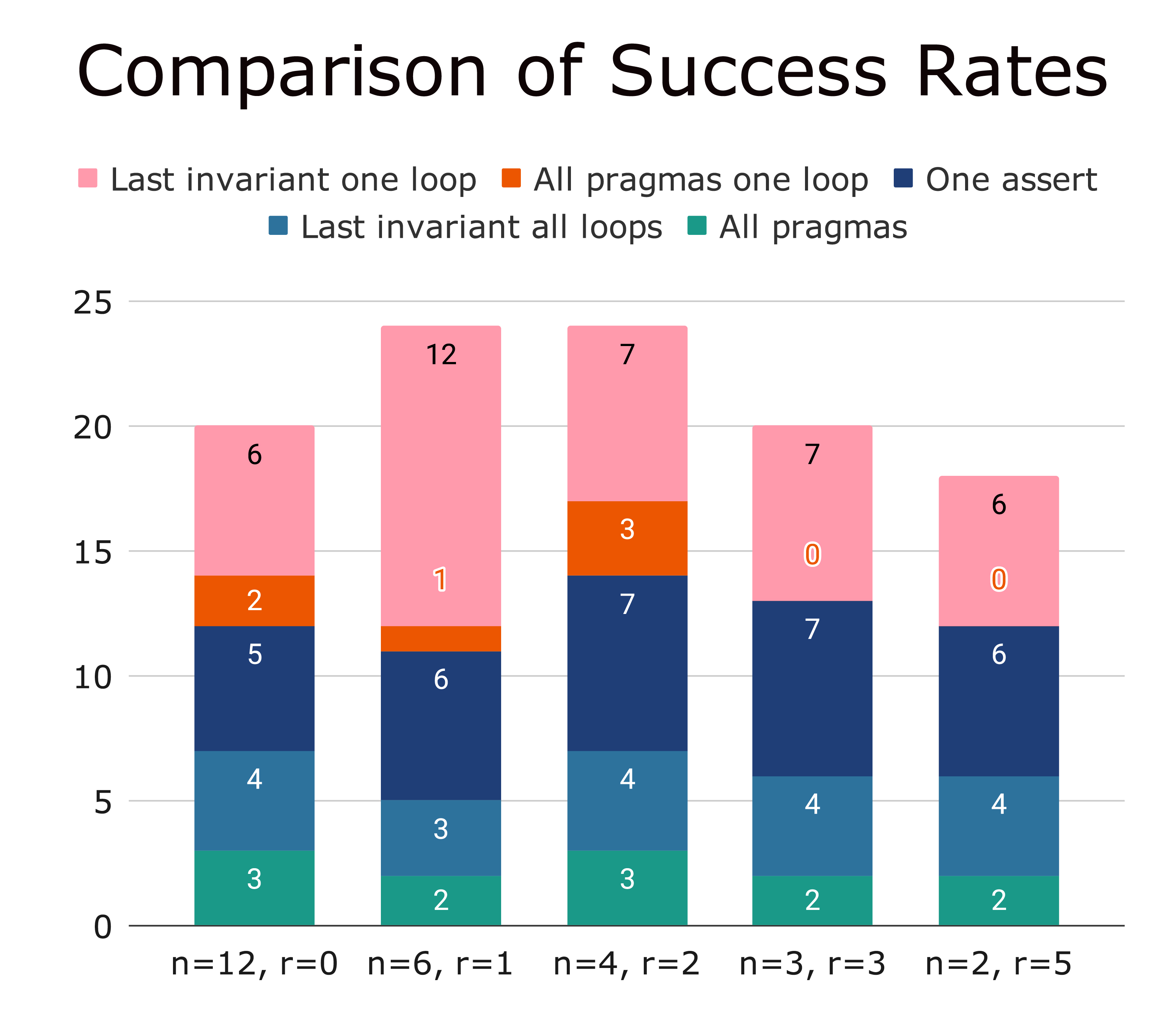}
	\caption[Results Graph - Comparison of Parameters]{Comparison of the number of programs solved per experiment. The bar is composed of the stacked number of solutions per benchmark.}
	\label{fig:experiment-comparison}
\end{figure}

Figure \ref{fig:experiment-comparison} is a graphical interpretation of Table \ref{tab:major-experiments}.

\begin{figure*}[t]
	\centering
	\vspace{-0.3cm}
	\includegraphics[width=0.9\textwidth]{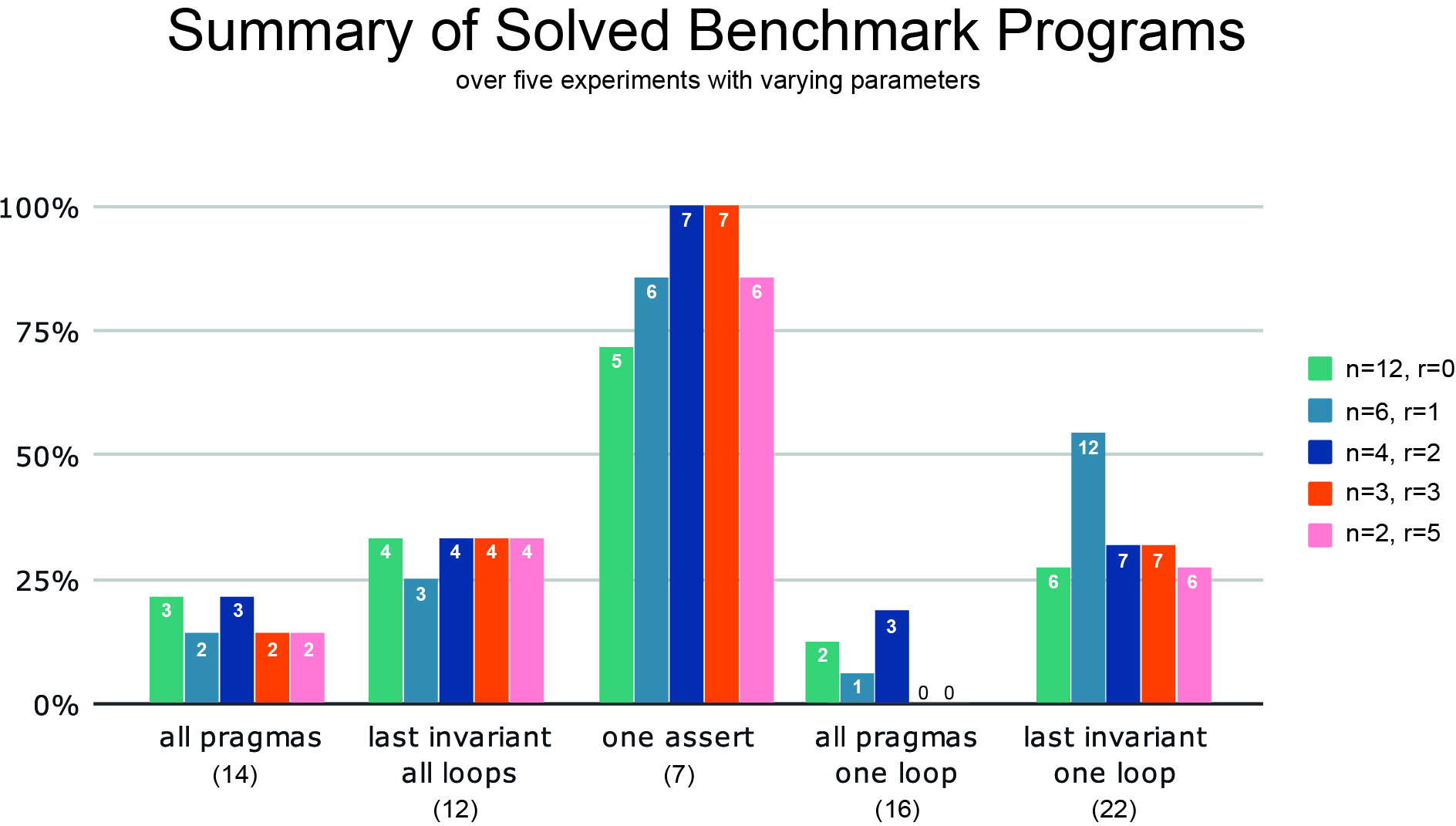}
	\caption[Results Graph - Total Success Rate]{Summary of the total amount of programs solved, per benchmark, across all five experiments. The chart is a 100\% bar chart, thus the y-axis represents the amount of programs that were solved, as a percentage of the total number of programs in the benchmark.}
	\label{fig:results-bargraph}
\end{figure*}

A further graphical representation of the data is Figure \ref{fig:results-bargraph}. The number of successfully solved programs, compared to the total number of programs in the benchmark, is depicted through a 100\% bar chart, to contrast performance on each of the benchmarks, per experiment.

\begin{table*}[ht]
\centering
\scriptsize
\caption{Breakdown of the number of programs solved in total, for each benchmark. The bottom row is the number of programs in the benchmark.}
\hspace*{-0.5cm}\begin{tabular}{lcccccc} 
\toprule
          & \multicolumn{1}{c}{All pragmas} & \multicolumn{1}{c}{Last invariant all loops} & \multicolumn{1}{c}{One assert} & \multicolumn{1}{c}{All pragmas one loop} & \multicolumn{1}{c}{Last invariant one loop} & \multicolumn{1}{c}{Sum}\\ 
\cmidrule(lr){2-2}\cmidrule(lr){3-3}\cmidrule(lr){4-4}\cmidrule(lr){5-5}\cmidrule(lr){6-6}\cmidrule(lr){7-7}
Total solved & 5 & 6 & 7 & 4 & 14 & 36\\
Total in benchmark & 14	& 12 & 7 & 16 & 22 & 71\\
\bottomrule
\end{tabular}
\label{tab:benchmark-totals}
\end{table*}

\begin{figure}[ht]
	\centering
	\includegraphics[width=0.5\textwidth]{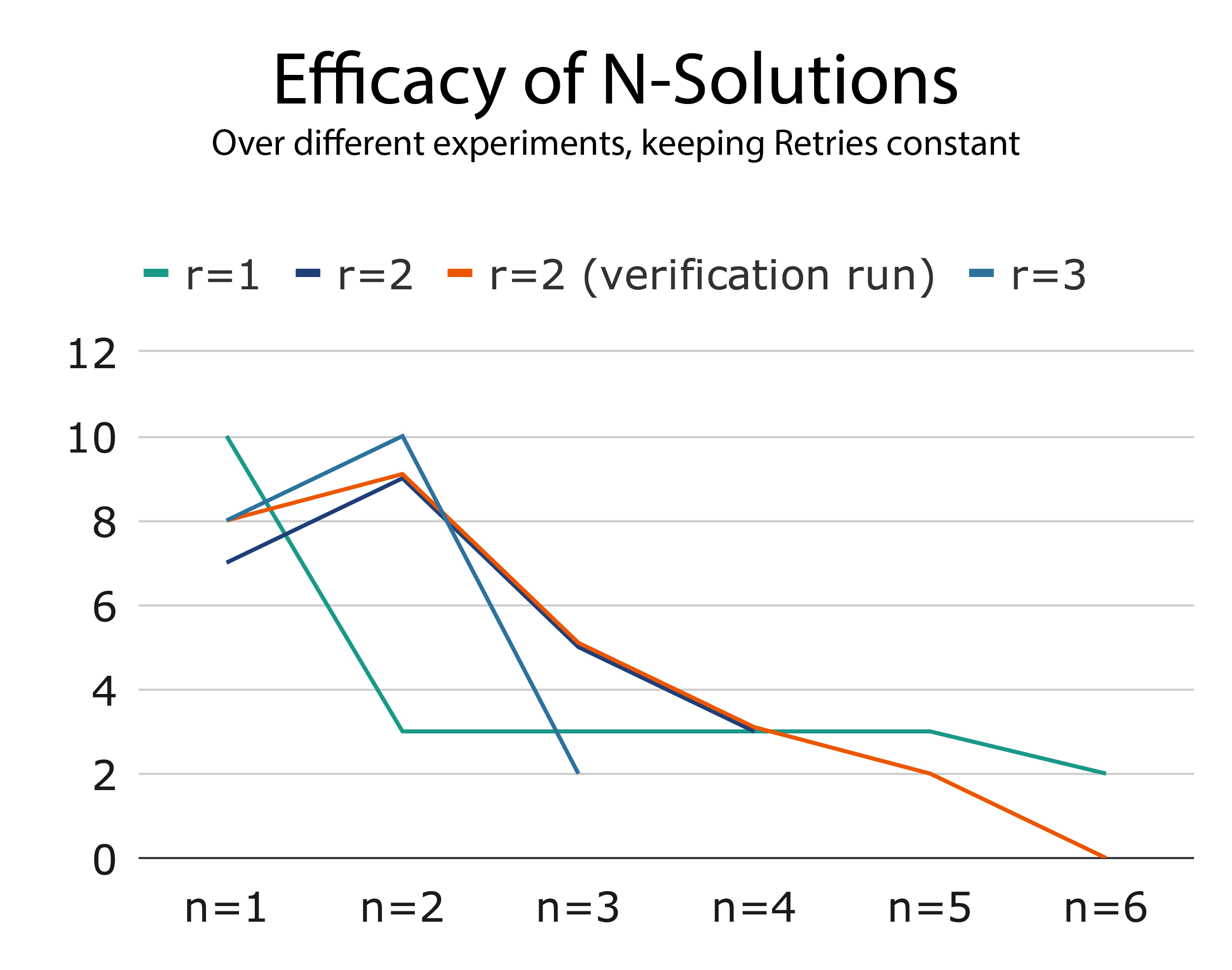}
	\caption[Line Graph - Efficacy of N-Solutions]{An overview of the efficacy of N-Solutions, between n=1 and n=6. The retries parameter is kept constant}
	\label{fig:efficacy-n-solutions}
\end{figure}

\begin{figure}[ht]
	\centering
	\includegraphics[width=0.5\textwidth]{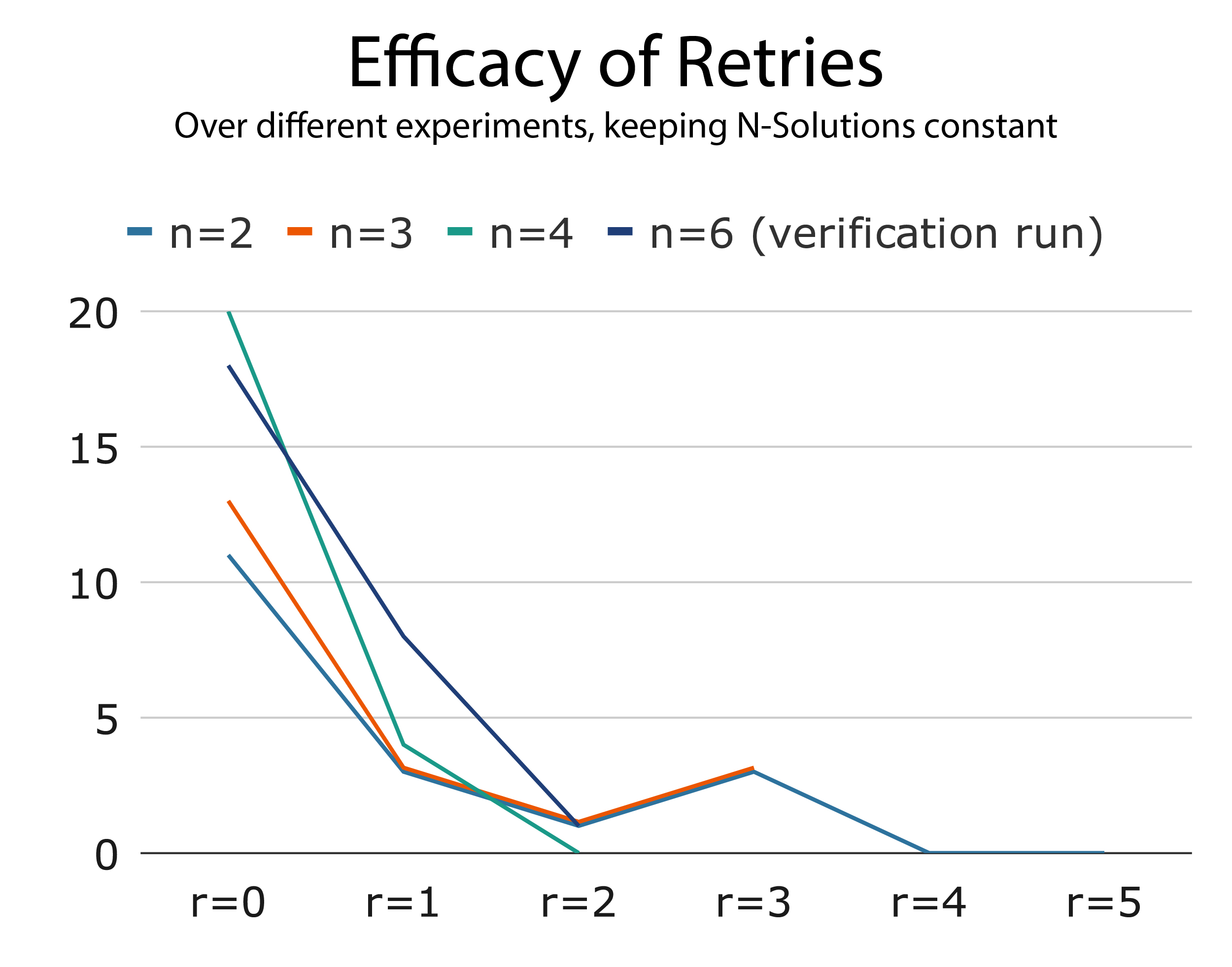}
	\caption[Line Graph - Efficacy of successive retries]{An overview of the efficacy of successive retries, between r=0 and r=5 (N-Solution parameter constant).}
	\label{fig:efficacy-retries}
\end{figure}

Figures \ref{fig:efficacy-n-solutions} and \ref{fig:efficacy-retries} contrast the effectiveness of successive retries and N-Solutions. Important to note here is that while one parameter is varied, the other is kept constant. In essence, each of the selected experiments was analyzed and the number of successful solutions was counted at each step, keeping one of the two parameters constant.

\subsection{Argu Results}
The programs that originate from Argu (see Section~\ref{subsec:programs}) are of high significance, as the programs cannot be part of GPT-4o's training data\footnote{If we take the statistics from Open AI's website to be up-to-date, at the time of writing, and provided the information is reliable.}. Thus, this makes it a notable benchmark, as the solutions could not have been learned, but had to be produced by the LLM without having seen the full program during pre-training.

Along with the programs being completely new for the LLM, the subject matter of the programs is also niche. Abstract argumentation theory is likely not among the major fields of academic research, and literature surrounding this topic is likewise uncommon. This makes the programs from the Argu repository potentially the most challenging in the benchmark.
With these premises in place, the results achieved by Marmaragan are surprising. Argu programs comprised just over half of the programs in the benchmark, totaling 36 out of 71. Out of these 36 programs, 16 were solved.

\subsection{Verification of Results}
Despite additional API costs, a replication of the two most successful experiments was conducted to attempt to show reproducibility.
The two most successful runs were $n = 6$, $r = 1$ and $n = 4$, $r = 2$, each reaching 24/71 solutions. These two were rerun, in order to check reproducibility. The results from this rerun were more successful than the initial runs, leading to a total of 25/71 successfully solved programs for $n = 4$, $r = 2$ and 26/71 for $n = 6$, $r = 1$.

\subsection{Chain of Thought Experiment}
The Chain-of-Thought experiment refers to a further experiment conducted, which differentiates itself from the main experiment in its prompting strategy. See section \ref{subsec:natural-language-prompt} for an explanation.

Using the set of parameters with the highest rate of successfully generated solutions: $n = 6$, $r = 1$, the experiment was conducted. In total, 25/71 programs were solved, equaling the average of the success rates from the original experiment and the verification run.


\section{\uppercase{Discussion}}\label{sec:discussion}
This section analyses the results of the experiments conducted with Marmaragan.

\subsection{General observations}

Experiments on the benchmark demonstrate Marmaragan's competence in generating annotations, both in the case of \texttt{Assert} statements and in the case of loop invariants, with a higher level of competence for \texttt{Assert} statements than for loop invariants. Overall, it generated correct annotations for 36 out of 71 (50.7\%) of the benchmark cases. These results highlight the potential for integrating formal verification into AI-assisted development, paving the way for safer and more reliable AI-generated software.

\subsection{N-Solutions and Retries Parameters}
Increasing the N-Solutions parameter, which determines the number of initial solution attempts, generally led to improved success rates in solving benchmark problems. The retry mechanism, which allows Marmaragan to attempt corrections based on error feedback, also proved to be an effective strategy. In many cases, experiments that incorporated retries outperformed those that relied solely on generating new solutions. This suggests that the model can effectively use error information to refine its approach.

The experiments reveal a complex interplay between N-Solutions and retries. The combinations of $n=6, r=1$ and $n=4, r=2$ yielded the best results, solving 24 out of 71 programs each. This suggests that a balance between initial solution attempts and correction opportunities is more effective than relying on either approach alone.  

As can be seen in Figures~\ref{fig:efficacy-n-solutions} and \ref{fig:efficacy-retries}, higher values of the \textit{$N$-Solutions} end \textit{Retries} parameters have deminishing returns. This is not surprising, as the model exhausts its most promising approaches in the first few attempts, with subsequent attempts becoming less likely to yield new solutions.



\section{\uppercase{Related Work}}\label{sec:related}

We are not aware of any work that brings LLMs and formal software verification together in the same way as we have proposed in this paper. But there is related work on leveraging LLMs for theorem proving and autoformalization in mathematics, from which valuable insights can be drawn for applying these techniques to the verification of software. This section reviews key works that inform our approach. These studies provide context for the capabilities and limitations of LLMs in formal reasoning, which Marmaragan seeks to extend to software verification.

Thor~\cite{thor} combines language models (LMs) with automated theorem proving systems using hammers to improve formal proof generation. Hammers are tools that bridge the gap between Interactive Theorem Provers (ITPs), which assist users in formulating proofs, and Automated Theorem Provers (ATPs), which independently verify conjectures. They achieve this by translating proof steps into ATP-compatible formats, selecting relevant premises, and integrating the ATP's solutions back into the ITP, enabling automated reasoning for individual proof steps~\cite{hammers-definition}. Sledgehammer~\cite{Sledgehammer} is a hammer for the Isabelle~\cite{Isabelle} ITP. Thor functions as follows: given a theorem and assumptions, it proves the conjecture by first allowing the LM to devise the proof steps, then appending $<hammer>$ annotations to individual sections, allowing Sledgehammer to complete the rest of the proof.  Tested on the PISA and MiniF2F benchmarks, Thor demonstrated higher success rates than individual components like Sledgehammer, solving 57\% of PISA benchmark problems. Despite its innovative approach, subsequent methods like Baldur have surpassed Thor's results, utilizing newer LLM technologies.

Baldur~\cite{baldur} generates entire formal proofs from theorem statements using LLMs and features a proof-repair mechanism that utilizes error messages to refine failed proofs. Unlike step-by-step proof generation, Baldur constructs full proofs and assesses their validity. It achieves a 47.9\% success rate on the PISA benchmark and demonstrates the effectiveness of proof repair and additional context for improving performance. Combining Baldur with Thor enhances results, solving 65.7\% of PISA problems, showcasing complementary strengths.

The Draft-Sketch-Prove (DSP) approach~\cite{draft-sketch-and-prove} addresses autoformalization by transforming informal mathematical proofs into verified formal proofs through three steps: drafting informal proofs, generating formal sketches, and completing gaps using automated provers. Using human-written and LLM-generated informal proofs, DSP achieves state-of-the-art performance on the MiniF2F benchmark, solving 42.6\% and 40.6\% of validation problems, respectively. Ablation studies highlight the importance of comments, multi-step reasoning, and integrating ATP tools like Sledgehammer.

Magnushammer~\cite{magnushammer} uses a transformer-based architecture to address premise selection, bypassing the need for extensive engineering. By training on premise selection datasets, it combines SELECT and RERANK algorithms to embed proof states and premises into a shared latent space, enabling relevance scoring. Magnushammer achieves state-of-the-art results on PISA, solving 59.5\% of problems, and boosts Thor's performance to a 71\% success rate when replacing Sledgehammer as the premise selection tool.

\section{\uppercase{Future Work}}\label{sec:future-work}

Marmaragan demonstrates the feasibility of an AI-powered annotation generator for SPARK 2014, but there remain several opportunities for further research and development. 

One potential direction is enabling the generation of pre- and postconditions by the LLM itself. Developers could define contracts for high-level functions, while the AI refines contracts for the invoked lower-level functions to preserve program correctness.

Testing Marmaragan on a benchmark of 16 programs yielded initial results, but validating its robustness and generalizability requires a larger dataset. Future benchmarks should include diverse SPARK programs spanning various domains and complexities.

Long-term goals include evolving Marmaragan into an industrial-grade tool, akin to how Copilot integrates into IDEs, by providing real-time annotation suggestions during SPARK code development.

Finally, an exciting avenue for long-term research based on the ideas in this paper would be to explore the possibility of employing LLMs to generate entire formally verified programs based on a conversation between a human project manager and an LLM about the intended behavior of the program.

\section{\uppercase{Conclusion}}\label{sec:conclusion}

This paper introduced Marmaragan, a proof-of-concept tool for generating SPARK 2014 annotations using LLMs. It integrates GNATprove to check whether the annotations complete the verification of the code. Marmaragan is thus a hybrid AI system that combines the power of LLMs with the trustworthiness of logic-based reasoning tools. Key techniques in the implementation of Marmaragan include utilizing pre-compiled GNATprove mediums, generating multiple solutions, retrying with additional context, and optional chain-of-thought prompting.

Benchmarking on 16 curated SPARK programs demonstrated Marmaragan's capabilities, particularly in generating \texttt{Assert} statements. The tool was able to solve 36 out of 71 benchmark cases.

This work highlights the potential for integrating formal verification into AI-assisted development, paving the way for safer and more reliable AI-generated software.

\section*{\uppercase{Acknowledgements}}

We would like to acknowledge the work done by Emmanuel Debanne. Based on ideas of the first co-author of this paper, Emmanuel developed Sparkilo, a closed-source tool similar to Marmaragan for generating annotations for SPARK 2014 programs, including features such as retries and chain of thought prompting. Sparkilo provided a great foundation for the ideas presented in this paper and laid out the groundwork for many of the features implemented in Marmaragan. Two functions authored by Emmanuel were included in Marmaragan.

Additionally, we would like to acknowledge the support of Tobias Philipp from \textit{secunet Security Networks AG}, who supported the two co-authors of this paper in the development of Marmaragan through his expertise in SPARK~2014.

\bibliographystyle{apalike}
{\small
\bibliography{bibliography}}

\end{document}